\documentclass[twocolumn,aps,prb,floats,showpacs,superscriptaddress]{revtex4}
\usepackage{amsmath}

\newcommand{\be}{\begin{equation}}
\newcommand{\ee}{\end{equation}}

\usepackage{epsfig}
\usepackage{graphicx}
\usepackage{bm}
\usepackage{amssymb}%

\def\gapp{\lower.35em\hbox{$\stackrel{\textstyle>}{\sim}$}}
\def\lapp{\lower.35em\hbox{$\stackrel{\textstyle<}{\sim}$}}

\begin{document}

\bibliographystyle{apsrev}

\title{Unconventional electromagnetic mode in neutral Weyl Semimetals}

\author{Yago Ferreiros}
\email{yago.ferreiros@csic.es}
\author{Alberto Cortijo}
\email{alberto.cortijo@csic.es}

\affiliation{Instituto de Ciencia de Materiales de Madrid, CSIC, Cantoblanco, 28049 Madrid, Spain.}

\begin{abstract}
We study light propagation in a neutral Weyl semimetal with the Fermi level lying at the Weyl nodes in the weak self-interacting regime. The nontrivial topology induces a screening effect in one of the two transverse gauge fields, for which we find two branches of attenuated collective excitations. In addition to the known topologically gaped photon mode, a novel massless and slightly damped excitation appears. Strikingly, at low energies this new excitation has a linear dispersion and it propagates with the same velocity than the electrons, while at energies well above the electron-hole continuum threshold it behaves as a massive attenuated photon with velocity similar to the speed of light in the material. There is a crossover at certain momentum in the direction perpendicular to the separation of the Weyl nodes above which the novel gapless mode enters into an overdamped regime. Regarding the unscreened gauge field we show that it is also attenuated, which is a non-topological property shared by Dirac semimetals as well. 
\end{abstract}
\pacs{78.20.-e, 03.65.Vf, 41.20.Jb}

\maketitle

\section{Introduction} 
Weyl semimetals (WSM) are three dimensional topological electronic systems whose low energy band structure is described by pairs of Weyl fermions separated in momentum space. They were first proposed to occur in a class of iridate materials\cite{WTVS11}, but it was only recently that they were experimentally realized\cite{XBA15,LWF15,LHG15,SAW15,XDZ15}. The topological properties of WSMs are encoded in a Chern-Simons (CS) term that stems from the axial anomaly. The anomaly is responsible of a plethora of phenomena like the axial magnetic effect\cite{FKW08,ZB12,CCG14}, chiral separation effect\cite{SZ04}, chiral magnetic waves\cite{KDY11} and, more recently, a new mechanism for the phonon Hall viscosity\cite{CFL15}. 

The experimental evidence of the WSM phase is based on ARPES and transport experiments\cite{XBA15,LWF15}, and on optical conductivity experiments\cite{XDZ15} (optical measurements have been done in Dirac semimetals as well\cite{XKK15,ZZL15}). The CS term induces birefringence\cite{CFJ90,G12}, and circular dichroism\cite{HQ15}. It has been also proposed that chiral electromagnetic waves exist and propagate along domain walls in magnetic WSMs\cite{ZZ14}. The interaction of light with magnons in WSMs has been addressed in\cite{HZRL14}, while plasmons in Weyl metals have also been studied\cite{LZ13,PBP14,ZCX15}. More recently, a new type of Helicon mode under applied magnetic fields has been proposed\cite{PKP15} at finite doping. Here we address the problem of light propagation in a WSM  at the neutrality point (i.e., the Fermi level lies at the Weyl nodes) by computing the poles of the full photon propagator. A similar approach was done for Quantum Electrodynamics (QED) under an external magnetic field\cite{F11}. For the case of the solid state-realized WSM there are two key aspects that are central: 1) nontrivial topology is present even in the absence of an external magnetic field and 2) contrary to QED there are two velocity scales in WSMs: the Fermi velocity of the electrons and the velocity of light in the material. We will see how these two features combine so that novel collective excitation behavior is realized. 
We will resort ourselves on a RPA treatment of the problem. This implies that we will effectively consider a non-self interacting Weyl fermions interacting with an external electromagnetic field. The validity of a RPA treatment has been extensively studied in the literature for two dimensional Dirac materials\cite{KUP12} and more recently in the general case of interacting three dimensional Dirac systems\cite{HJS15,THJ15}. Following these references, some of the results presented here might be extended to the self-interacting case under particular conditions ($1/N$ expansions and so).

The rest of the paper is structured as follows. In section II we describe the model for WSM that we use and set notations. In section III, we give the neccessary details for the computation of the polarization function and explain it. In section IV, we compute the spectrum of the electromagnetic collective modes through the spectral function, and describe the most salient features observed. In section V, we discuss the results and mention some future lines to follow from this work.
\section{The model} 
Our starting point is the low energy action of a WSM coupled to an electromagnetic field:
\begin{equation}
S=\int d^4x\overline{\Psi}\Big(i\gamma^0 (\partial_0+ieA_0)-iv\gamma^i (\partial_i+\frac{ie}{c}A_i)+b_i\gamma^i\gamma^5\Big)\Psi,
\label{clasaction}
\end{equation}
where $v$ is the Fermi velocity and $c$ is the speed of light in the corresponding material. This action describes two Weyl fermions separated in momentum space by a vector $2\bm{b}/v$ so that time reversal symmetry is broken. Without loss of generality will fix $\bm{b}=b\hat{x}_3$. Quantum corrections to (\ref{clasaction}) are well known to generate a nonzero odd part of the polarization function of the photon in the form of a CS term \cite{JK99,PV99,CP99,AGS02,C99}. The coefficient accompanying the CS term is ill defined, as it depends on the routing of internal momenta in the loop integrals as well as on the regularization method. In a solid state realization of a WSM this coefficient can be fixed by means of the bulk-boundary correspondence relating the theory in the bulk with the Fermi arcs at the edges\cite{GT13,RH14}. Another way is by considering the 3D band structure of a WSM as a set of independent 2D band structures, each of them with a well defined Chern number\cite{BB11,KYY11,G12}.

In addition to the action for the fermionic sector (\ref{clasaction}), the action for the electromagnetic field is:
\begin{equation}
S=\int d^4x\frac{1}{2}\Big(\epsilon |\bm{E}|^2-\frac{|\bm{B}|^2}{\mu}\Big),
\label{actionphoton}
\end{equation}
where $\epsilon=\epsilon_{0}\epsilon_r$ and $\mu=\mu_0\mu_r$ are the dielectric permittivity and magnetic permeability of the material respectively. The electric and magnetic fields are related to the gauge field $A_\mu$ as:
\begin{equation}
E_i=\frac{1}{c}\partial_0A_i-\partial_iA_0,\quad B_i=\frac{1}{c}\epsilon_{ijk}\partial^jA^k.
\label{actionphoton2}
\end{equation}

\section{Polarization function} 
In what follows we will obtain the 1-loop order polarization function in terms of $A_\mu$. The computation has been performed and deeply examined by several authors, both the parity-even\cite{A041,A06,BCT08,A042} and parity-odd\cite{JK99,PV99,CP99,AGS02,C99} parts of the polarization function. We can write it as:
\begin{equation}
\Pi^{\mu\nu}=\Pi^{\mu\nu}_0+\Pi^{\mu\nu}_b,
\label{polarization}
\end{equation}
where $\Pi^{\mu\nu}_0$ is the contribution at zero order in $b$, and $\Pi^{\mu\nu}_b$ is the first order correction. In\cite{A041} it was shown that the polarization function is analytic in $b$, and the correction to the even part is second order in $b$, breaks gauge invariance and gives a mass to the photon. Imposing gauge invariance by using a regularization which preserves it, as dimensional or Pauli-Villars regularizations, one obtains a zero correction\cite{A041,A042}. Regarding the odd part, it is known to be linear in $b$\cite{C99,PV99} and no corrections appear beyond the first order. Hence eq. (\ref{polarization}) is exact, i.e. there are not higher order corrections in $b$.

The computation of $\Pi^{\mu\nu}_0$ is standard. Using dimensional regularization and in the minimal subtraction renormalization scheme:
\begin{equation}
\Pi^{\mu\nu}_0=\frac{e^2}{4\pi v^3}\Big(\frac{v}{c}\Big)^{2-\delta_{\mu0}-\delta_{\nu0}}\Pi(p^2/\mu_0^2)(p^2\eta^{\mu\nu}-p^\mu p^\nu),
\label{polfunczeroorder}
\end{equation}
\begin{equation}
\Pi(p^2/\mu_0^2)=\frac{1}{9\pi}(3\log(-p^2/\mu_0^2)-5),
\label{polfunczeroorder2}
\end{equation}
where the four momentum is $p_\mu=(\omega,v\bm{p})$, and where the space components of $\Pi^{\mu\nu}_0$ have a relative weight $v/c$ with respect to the time components, as can be read from eq. (\ref{clasaction}). From now on we will redefine $\bm{p}$ as $v\bm{p}\rightarrow\bm{p}$ so that $p^2=\omega^2-|\bm{p}|^2$. The parameter $\mu_0$ is an energy scale that appears in the process of dimensional regularization (see for example \cite{S07}). The appearance of this parameter reflects the formal absence of any characteristic scale in the linear electronic spectrum. It could be fixed by either the experiment or by invoking a standard renormalization program (here we are dealing with a modification of standard quantum electrodynamics, that is well known to be renormalizable) or related to the lattice cutoff of the underlying bandstructure by considering a full lattice model of the Weyl semimetal, keeping in mind that there are several materials with very different lattice structures proposed to be Weyl semimetals. For typical values of lattice spacings of the order of \r{A}ngstr\"oms, $\mu_{0}$ can be estimated to be $\mu_0\sim 10$ eV, however, the reader should keep in mind that the low energy/long wavelength aspects of the results presented here do not depend on the particular value of $\mu_0$ as long as all the frequencies and momenta involved in the problem are presented in a \emph{dimensionless} fashion (that is, divided by $\mu_{0}$) no matter which regularization scheme is used.  

Note that  the function $\Pi(p^2/\mu_0^2)$ in eq. (\ref{polfunczeroorder2}) has a constant imaginary part for $p^2>0$ which defines the electron-hole continuum threshold. A normal procedure would be to take the local limit $\omega/|\bm{p}|\gg1$ in $\Pi^{\mu\nu}_0$. As we know for graphene, at the neutrality point and zero temperature no well defined plasmon mode exists within the random phase approximation (RPA). However, it was shown that a plasmon pole appears beyond RPA when the momentum dependence in the polarization tensor is maintained\cite{GFM08}, which highlights the importance of not considering the local limit in some circumstances. Here we will limit ourselves to the RPA approximation but we will keep the momentum dependence, and we will see that staying away from the local limit is crucial for capturing the physics near the electron-hole continuum threshold.

$\Pi^{\mu\nu}_b$ is precisely the CS term. We invoke the boundary-bulk correspondence to fix its value\cite{GT13,RH14}. On the $A_0=0$ gauge $\Pi^{\mu\nu}_b$ has only one independent component:
\begin{equation}
\Pi^{12}_b=-\Pi^{21}_b=i\frac{e^2b}{2\pi^2c^2v}\,\omega.
\label{oddpol}
\end{equation}
The presence of this non-diagonal term in the polarization tensor induces gyrotropy in the WSM, so that left- and right-handed polarized light propagate at different speeds\cite{CFJ90,G12,HQ15}.

Since we are working under the free electron approximation, eq.(\ref{oddpol}) is exact. In an interacting theory, $\Pi^{12}_{b}$ acquires dynamically generated corrections( see \cite{GMS13} for an explicit discussion when interactions at constant external magnetic field and finite densities are considered, or \cite{BPV15} for corrections in the strong coupling regime).

\section{Collective modes.} 

The inverse full photon propagator can be written as:
\begin{equation}
(\Delta^{\mu\nu})^{-1}=(\Delta^{\mu\nu}_{free})^{-1}-\Pi^{\mu\nu}_0-\Pi^{\mu\nu}_b,
\label{invs phot prop}
\end{equation}
where the inverse free propagator can be obtained from eqs. (\ref{actionphoton},\ref{actionphoton2}). To obtain the collective excitations we look for the poles of the full propagator. The equation $det[(\Delta^{\mu\nu})^{-1}]=0$ on the $A_0=0$ gauge reads:
\begin{widetext}
\begin{eqnarray}
&&\Big(1-g\Pi(p^2/\mu_0^2)\Big)\omega^2\,\times\,\Big(1-g\Pi(p^2/\mu_0^2)\Big)\Big(\omega^2-\beta\Gamma(p^2/\mu_0^2)|\bm{p}|^2\Big)\,\times\nonumber\\
&&\times\,\left(\Big(1-g\Pi(p^2/\mu_0^2)\Big)\Big(\omega^2-\beta\Gamma(p^2/\mu_0^2)|\bm{p}|^2\Big)-J\big(\frac{\omega^2}{\mu_0^2},\frac{|\bm{p}_\perp|^2}{\mu_0^2},\frac{p_3^2}{\mu_0^2}\big)\right)=0,
\label{eq det}
\end{eqnarray}
\end{widetext}
where $|\bm{p}_\perp|^2=p_1^2+p_2^2$ represents the momentum in the plane perpendicular to $\bm{b}$, $g=e^2/(4\pi v\epsilon_{0}\epsilon_r)$ is the fine structure constant, $\beta=c^2/v^2$ gives the squared ratio between the speed of light and the Fermi velocity, and:
\begin{equation}
\Gamma(p^2/\mu_0^2)=\frac{1-\frac{g}{\beta}\Pi(p^2/\mu_0^2)}{1-g\Pi(p^2/\mu_0^2)},
\end{equation}
\begin{eqnarray}
J(x,y,z)&=&\frac{4g^2}{\pi^2}\frac{b^2}{1-g\Pi(x-y-z)}\times\nonumber\\
&\times &\frac{x-\beta\Gamma(x-y-z)y}{x-\beta\Gamma(x-y-z)\big(y+z\big)}.
\end{eqnarray}
We will take the Fermi velocity of the WSM similar to that of graphene ($v=10^6m/s$) and assume a coupling constant $g=1$, which is reached with a relative permittivity of the material of $\epsilon_r\approx2.3$. The speed of light of the WSM is given by $c=c_{vacuum}/\sqrt{\epsilon_r}$ (assuming a relative permeability $\mu_r=1$) and the value of the squared ratio between the speed of light and the Fermi velocity is $\beta\approx3.9\times10^4$. Since we are not taking into account interactions between Weyl electrons, a value of the effective coupling $g=1$ is not problematic, and this precise value has been chosen for convenience. In realistic solid-state WSM, however, it is not expected to have these values for $v_{F}$ and/or larger values of $\epsilon_r$, resulting effectively in an smaller value of $g$.

\begin{figure}
\includegraphics[trim=34 0 0 -30, width=8cm,scale=0.18]{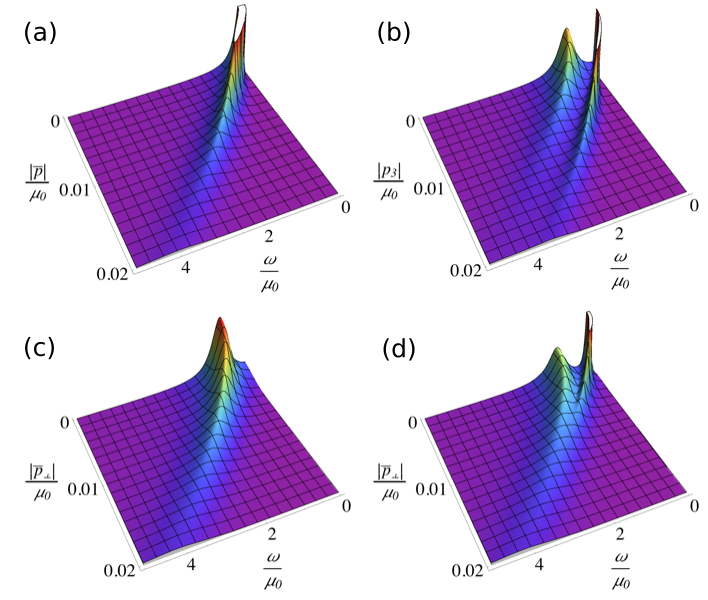}
\caption{(Color online) (a) Spectral function $\rho$ for the unscreened gauge field as a function of the frequency $\omega$ and momentum $|\bm{p}|$ in units of $\mu_0$. (b) Spectral function $\rho$ for the screened gauge field as a function of $\omega$ and $|p_3|$ in units of $\mu_0$, for zero transverse momentum $|\bm{p}_\perp|=0$. (c,d) Spectral function $\rho$ for the screened gauge field as a function of the frequency $\omega$ and the transverse momentum $|\bm{p}_\perp|$ in units of $\mu_0$, for (c) $|p_3|=0$ and (d) $|p_3|/\mu_0=3\times10^{-3}$. All plots for the screened gauge field are presented for $b^2/\mu_0^2=3$.}
\label{fig 1}
\end{figure}

There is a zero mode $\omega^2=0$ corresponding to the unphysical longitudinal gauge field, arising from the residual gauge freedom, as $A_0=0$ does not completely fix the gauge. For the remaining part, there are two physical transverse gauge fields. The dispersion relation of the unscreened field can be obtained from:
\begin{equation}
\Big(1-g\Pi(p^2/\mu_0^2)\Big)\Big(\omega^2-\beta\Gamma(p^2/\mu_0^2)|\bm{p}|^2\Big)=0.
\label{eq unscreened mode}
\end{equation}
Instead of directly solving the equation, we will define a spectral function for this field:
\begin{equation}
\rho=-2\Im\Bigg[\frac{\mu_0^2}{\big(1-g\Pi(p^2/\mu_0^2)\big)\big(\omega^2-\beta\Gamma(p^2/\mu_0^2)|\bm{p}|^2\big)}\Bigg].
\end{equation}
The spectral function is plotted in Fig. \ref{fig 1}(a). We see that there is a peak with finite width, which corresponds to a damped mode, so that the delta function corresponding to a free photon is replaced in a WSM by an attenuated excitation. Hence our first finding is that light is attenuated when propagating through a WSM. This behavior is rooted to the existence of two different velocity scales: the value of the Fermi velocity is much smaller than the speed of light. This mismatch is encoded in the function $\Gamma$ appearing in eq. (\ref{eq unscreened mode}), which fulfills $\lim_{\beta\rightarrow1}\Gamma=1$. This means that in the limit where the value of the Fermi velocity matches the value of the speed of light in the material, we have a non attenuated free photon traveling at the speed of light with dispersion relation $\omega=|\bm{p}|$. Although the gapless nature of WSMs means that there is no gap to overcome for electron-hole pair creation, we just showed that this is not enough for light attenuation, and that damping is generated for a system of gapless electrons only when there exist two non-equal competing velocity scales. This happens also in Dirac semimetals since this property is not related to the CS term. 

\begin{figure*}
\textbf{(a)}
\begin{minipage}{.46\linewidth}
\includegraphics[scale=0.175]{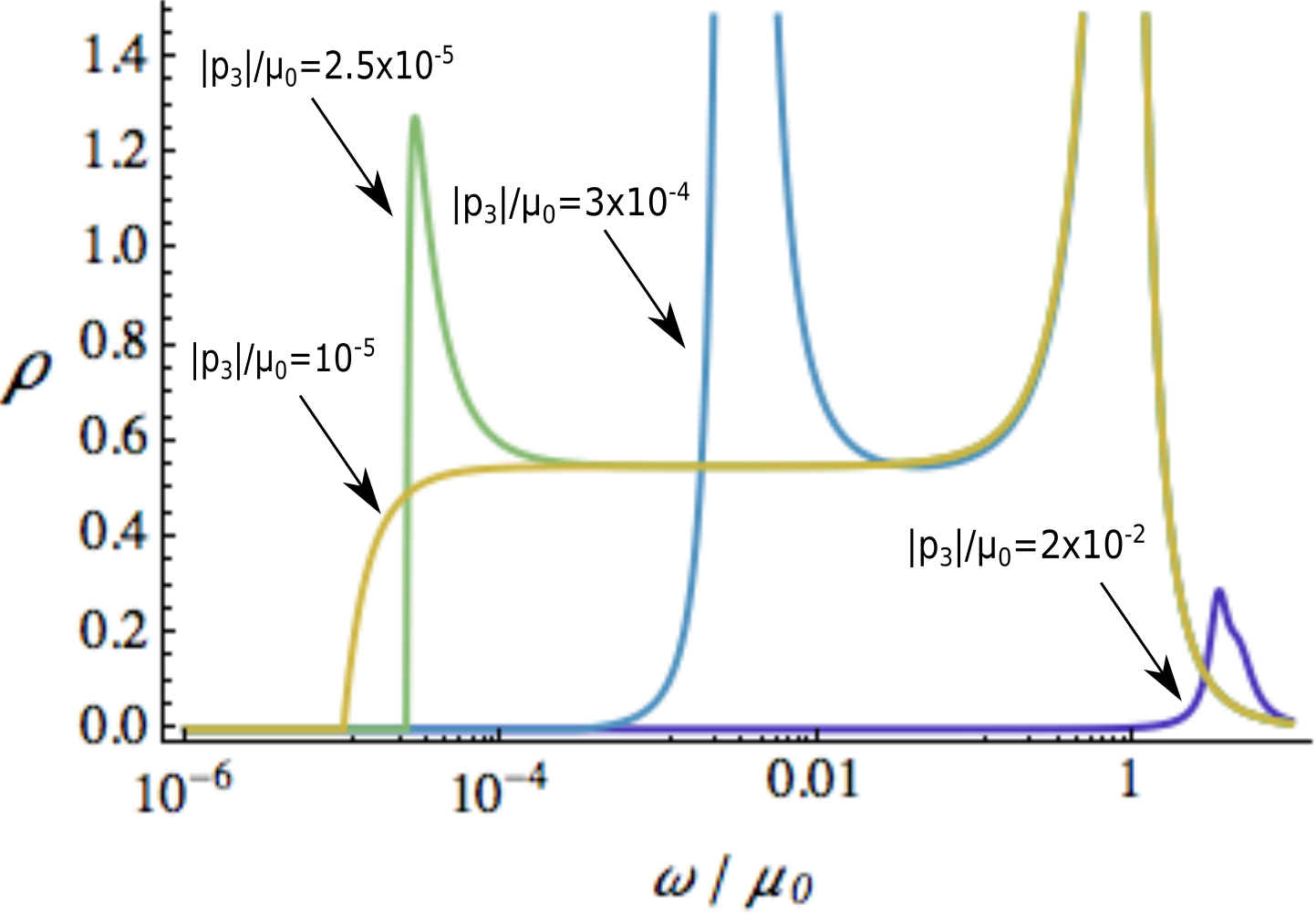}
\end{minipage}
\textbf{(b)}
\begin{minipage}{.46\linewidth}
\includegraphics[scale=0.20]{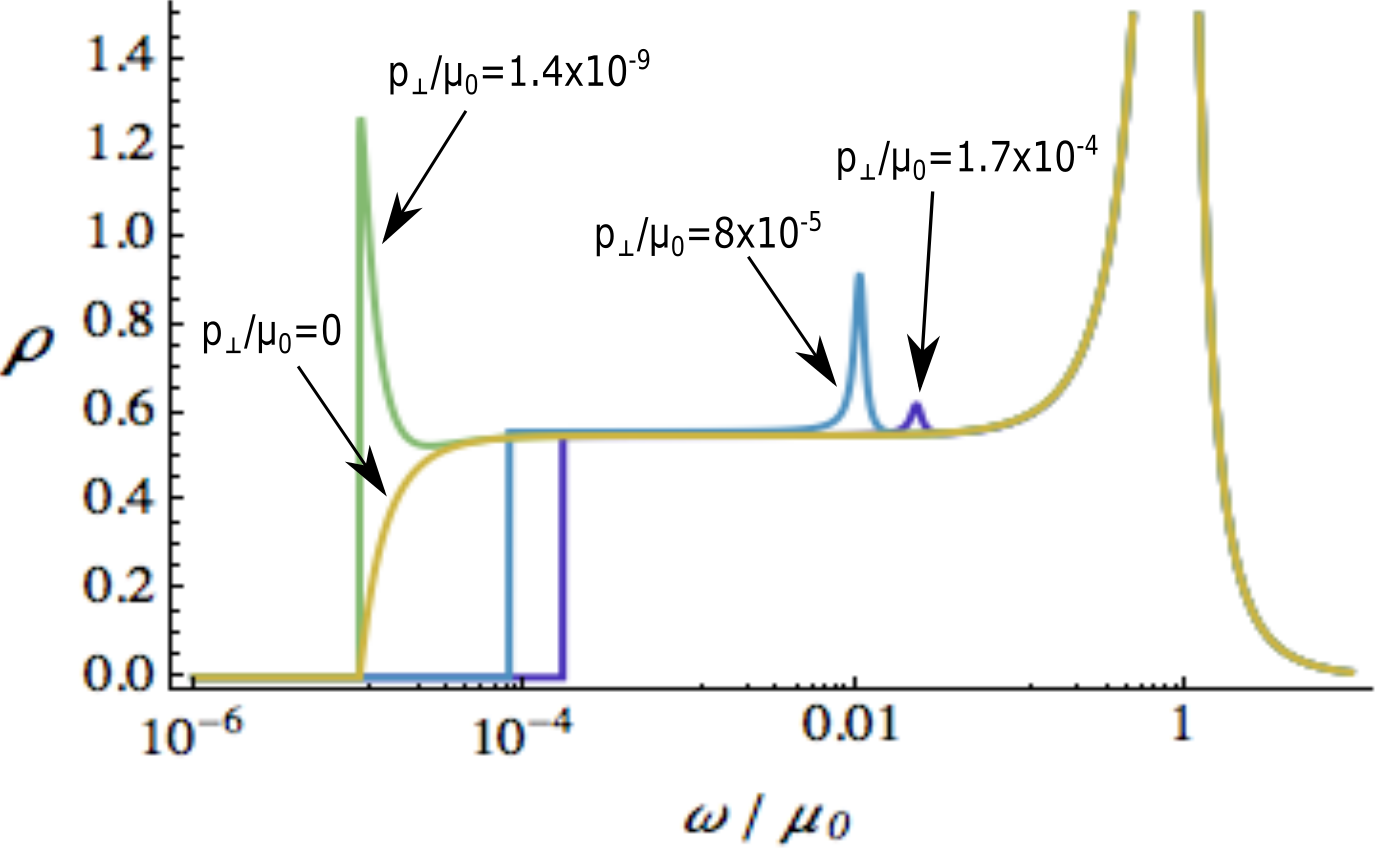}
\end{minipage}
\caption{(Color online) Both plots are presented for a value $b^2/\mu_0^2=3$ and in a logarithmic scale for the horizontal axis. (a) Spectral function $\rho$ for the screened gauge field as a function of the frequency $\omega$ in units of $\mu_0$ for $\bm{p}_\perp=0$ and different values of $|p_3|$. (b) Spectral function $\rho$ for the screened gauge field as a function of the frequency $\omega$ in units of $\mu_0$ for $p_3=0$ and different values of $|\bm{p}_\perp|$.}
\label{fig 2}
\end{figure*}

The nontrivial topology of the WSM phase shows up inducing a screening in the other transverse gauge field. The dispersion relation can be obtained from the second line of eq. (\ref{eq det}). Instead of solving the equation, we will again introduce a spectral function for this field:
\begin{eqnarray}
\rho &=&-2\Im\Bigg[\mu_0^2\Big\{\big(1-g\Pi(p^2/\mu_0^2)\big)\big(\omega^2-\beta\Gamma(p^2/\mu_0^2)|\bm{p}|^2\big)-\nonumber\\
&-&J\big(\frac{\omega^2}{\mu_0^2},\frac{|\bm{p}_\perp|^2}{\mu_0^2},\frac{p_3^2}{\mu_0^2}\big)\Big\}^{-1}\Bigg].
\end{eqnarray}
In the analysis following we will consider two separate cases: $\bm{p}_\perp=0$ and $\bm{p}_\perp\neq0$.

i) $\bm{p}_\perp=0$. In Fig. \ref{fig 1}(b) the spectral function is plotted as a function of $\omega$ and the momentum $|p_3|$ in the direction of $\bm{b}$. Two branches of attenuated collective excitations coexist, one of them clearly gaped. The gaped nature of the screened field is expected in WSMs, as the CS term acts as a topological mass for the photon\cite{CFJ90}. However we see that the second branch corresponds to a new excitation which seems to be gapless and considerably less attenuated than its gaped companion. In fact, at a certain large momentum $p_3$ there is a crossover where the gaped excitation enters into an overdamped regime, as can be seen from the absence of a well defined peak in Fig. \ref{fig 1}(b). To highlight the behavior of the novel gapless excitation at low frequencies near the electron-hole continuum threshold $\omega=|\bm{p}|$, we plot in Fig. \ref{fig 2}(a) the spectral function as a function of the frequency in a logarithmic scale, for different values of $p_3$. We see that there is a low energy crossover also for the novel excitation from an underdamped to an overdamped regime as the wavelength increases. Overdamping appears for frequencies near the electron-hole continuum threshold. This suggests a gapless nature of this new excitation with dispersion relation $\omega\sim|\bm{p}|$ at low energies. We see also that increasing the momentum $p_3$, the gapless branch deviates from the linear dispersion behavior and approaches the gaped one. From Fig. \ref{fig 1}(b) it is clear that the gapless branch starts to behave as a massive excitation for large enough momentum $p_3$.

ii) $\bm{p}_\perp\neq0$. Figs. \ref{fig 1}(c,d) show the spectral function as a function of $\omega$ and $|\bm{p}_\perp|$ for the values $p_3=0$ and $|p_3|/\mu_0=3\times10^{-3}$ respectively. It can be seen that increasing the perpendicular momentum produces a merging of the two branches into a single peak. This is a result of a second, high energy crossover for the gapless excitation which again becomes overdamped, but now at big enough $\bm{p}_\perp$. For energies near the electron-hole continuum threshold, we can see in Fig. \ref{fig 2}(b) that small fluctuations in $\bm{p}_\perp$ drive the gapless excitation across the low energy crossover from the overdamped to the underdamped regime, so in a realistic situation when the propagation is not perfectly aligned with the $\bm{b}$ direction the gapless branch behaves as a well defined attenuated excitation at low energies.

It is important to note that the behavior near the electron-hole continuum would have never been captured in the local limit $\omega/|\bm{p}|\gg 1$ of the polarization function. In particular the low energy crossover into the overdamped regime of the massless branch would have never been captured without considering the full momentum dependence.

\section{Discussion and conclussions} 

We have studied light propagation in a neutral WSM where the Fermi level lies at the Weyl nodes. In this situation there are two degrees of freedom corresponding to two transverse gauge fields, while the plasmon longitudinal mode that appears at finite chemical potential is absent. Besides the generation of a plasmon mode, the presence of a plasma frequency would also induce a gap to both transverse gauge fields, however at the neutrality point there is only one gaped mode, whose mass is a consequence of the nontrivial topology.  

Our first finding is that light is attenuated when propagating through both Weyl and Dirac semimetals. In the case of WSMs, nontrivial topology shows up as a screening effect in one of the two transverse gauge fields, for which we find two branches of attenuated collective excitations. In addition to the known topologically gaped photon mode, a novel massless attenuated excitation appears, which is our second and most important finding. This splitting of a transverse field into two branches was obtained before for QED in the presence of an external magnetic filed\cite{F11}, where the ciclotron frequency defines the mass threshold for the electron-hole continuum. In the case of a WSM there is no mass threshold as nontrivial topology is present in the absence of magnetic fields. Furthermore for the case of the solid state-realized WSM the Fermi velocity of the electrons is much smaller than the speed of light, which translates into a completely novel collective behavior of the two branches. 

In particular we find that in a realistic situation of light collimated along the direction of $\bm{b}$ but with small deviations in the perpendicular direction, the massless branch shows up as a well defined and damped peak in the spectral function. At low energies its dispersion relation is given by $\omega\simeq v|\bm{p}|$, which defines an excitation traveling coherently with the electrons at the Fermi velocity. At the other limit of the spectrum ($\omega\gg v|\bm{p}|$) the gapless branch behaves as a massive damped photon with velocity close to the speed of light. In the situation where light propagation is not aligned with $\bm{b}$ and the momentum in the transverse direction is large enough, a crossover exists and the massless branch enters into an overdamped regime, corresponding to a peak with divergent width in the spectral function. 

The novel optical excitation could provide a new and clean signature for WSMs. It could be detected in experiments as an additional resonance on the transmission coefficient of light. There are also indirect proofs, like electron energy-loss spectroscopy (EELS) or subluminal Cherenkov radiation spectroscopy\cite{SWK01} stemming from the small phase velocity of the novel excitation. In addition, due to the matching of the propagation velocity of the novel mode with the Fermi velocity of the electrons, a counter intuitive enhancement of the effective dressed electron-photon interaction is expected. This is qualitatively different from the effect of the Coulomb interaction enhancement due to the renormalization of the Fermi velocity that appears, for instance, in graphene\cite{GGV94}. It should be noted that, although their effect is negligible on the bulk propagation of light, the 1D Fermi arc chiral states at the edges of a WSM would give additional resonances as surface electromagnetic modes. However, the polarization tensor of 1D chiral electrons has already been computed in the literature\cite{JR85,FC14} and therefore the Fermi arc contribution can be distinguished from bulk propagation. 

\section{Acknowledgments} We thank M. A. H. Vozmediano for enlightening discussions and encouragement. This research is partially supported by the Spanish MECD Grant No. FIS2011-23713, the European Union structural funds and the Comunidad de Madrid MAD2D-CM Program (S2013/MIT-3007), and by the European Union Seventh Framework Programme under grant agreement no. 604391 Graphene Flagship.


\end{document}